\documentclass[a4paper,aps,twocolumn]{revtex4}
\RequirePackage[english]{babel}
\RequirePackage[latin1]{inputenc}
\RequirePackage[T1]{fontenc}
\RequirePackage{mathrsfs}
\RequirePackage{amsmath}
\RequirePackage{amssymb}
\RequirePackage{amsbsy}
\RequirePackage{color}
\RequirePackage{bm}
\begin{document}
\title{\bf{A geometrical assessment of spinorial energy conditions}}
\author{Luca Fabbri}
\affiliation{INFN \& dipartimento di fisica, universit{\`a} di Bologna,\\
via Irnerio 46, 40126 Bologna, ITALY}
\date{\today}
\begin{abstract}
We consider the problem of energy for spinor fields coupled to their surrounding curved-twisted space-time, and we show that when treated geometrically we cannot even be certain that there is a problem for the energy in the first place.
\end{abstract}
\maketitle
\section{Introduction}
In the usual development of field theory, the discussion about spinor fields starts by neglecting their surrounding gravitational field and then acknowledging that solutions of the Dirac equation with positive energy come together with complementary solutions of the same Dirac equation with negative energy: since there appears to be no matter with negative energy people were prompted to look for a way to solve this apparent problem, and the need for its solution was the single most important reason for which we ventured in the realm of field quantization. Nowadays, non-convergence of perturbative series and infinities in its terms, or merely the lack of the very definition of the field operators, keep pushing us beyond quantum field theory, and although it is obvious that quantum field theory must be completed or revised, nobody knows how to do it.

When trying to solve a problem, sometimes it is enough to go back to its roots, see where the issue lies and fix it, but other times it may be necessary to go even before, to assess whether there really is a problem in the first place, and if we do this now, we immediately recognize there is something odd in the fact that we discussed energy never accounting for gravity: since energy density is the source of the gravitational field, it is obvious that the usual ways of approaching the topic cannot be conclusive. Therefore, to stand a chance the energy must be investigated using its full coupling in the gravitational field equations.

However, because here matter is spinor fields, possessing spin beside energy, the full coupling must account for torsion as well as gravity, which is what we will do.
\section{Generally-coupled spinors}
In this work, we refer to \cite{Fabbri:2014dxa} for the basic notations and the fundamental assumptions: these are consistency conditions such as the requirement that all fields propagate at the least-order derivative, the causal structure and the continuity. When these are implemented, torsion can be taken to be the completely antisymmetric dual of an axial vector $W_{\alpha}$ beside the spin connection $\Omega^{a}_{\phantom{a}b\pi}$ and the gauge potential $A_{\mu}$ as usual, with curl $(\partial W)_{\alpha\nu}$ alongside to the Ricci tensor $R_{\mu\nu}$ and the Maxwell tensor $F_{\mu\nu}$ to define a dynamical Lagrangian for the geometric background.

In reference \cite{Fabbri:2016msm} we have summarized essential elements of the Clifford algebra, and here we will recall only those that are the most important, starting from the matrices
\begin{eqnarray}
&\{\boldsymbol{\gamma}_{a},\boldsymbol{\gamma}_{b}\}
=2\eta_{ab} \boldsymbol{\mathbb{I}}
\end{eqnarray}
from which we also define
\begin{eqnarray}
&\boldsymbol{\sigma}_{ab}
=\frac{1}{4}[\boldsymbol{\gamma}_{a},\boldsymbol{\gamma}_{b}]\\
&\boldsymbol{\sigma}_{ab}
=-\frac{i}{2}\varepsilon_{abcd}\boldsymbol{\pi}\boldsymbol{\sigma}^{cd}
\label{pi}
\end{eqnarray}
verifying the relationships
\begin{eqnarray}
&\boldsymbol{\gamma}_{a}\boldsymbol{\gamma}_{b}
\!=\!\eta_{ab} \boldsymbol{\mathbb{I}}\!+\!2\boldsymbol{\sigma}_{ab}\label{gamma2}\\
&\boldsymbol{\gamma}_{i}\boldsymbol{\gamma}_{j}\boldsymbol{\gamma}_{k}
=\boldsymbol{\gamma}_{i}\eta_{jk}-\boldsymbol{\gamma}_{j}\eta_{ik}+\boldsymbol{\gamma}_{k}\eta_{ij}
+i\varepsilon_{ijkq}\boldsymbol{\pi}\boldsymbol{\gamma}^{q}\label{gamma3}
\end{eqnarray}
and with the pair of conjugate spinors we define
\begin{eqnarray}
&2\overline{\psi}\boldsymbol{\sigma}^{ab}\boldsymbol{\pi}\psi\!=\!\Sigma^{ab}\label{pt}\\
&2i\overline{\psi}\boldsymbol{\sigma}^{ab}\psi\!=\!S^{ab}\label{t}\\
&\overline{\psi}\boldsymbol{\gamma}^{a}\boldsymbol{\pi}\psi\!=\!V^{a}\label{a}\\
&\overline{\psi}\boldsymbol{\gamma}^{a}\psi\!=\!U^{a}\label{v}\\
&i\overline{\psi}\boldsymbol{\pi}\psi\!=\!\Theta\label{p}\\
&\overline{\psi}\psi\!=\!\Phi\label{s}
\end{eqnarray}
such that
\begin{eqnarray}
&\Sigma^{ab}\!=\!-\frac{1}{2}\varepsilon^{abij}S_{ij}\\
&S_{ab}(\Phi^{2}\!+\!\Theta^{2})\!=\!\Phi U^{j}V^{k}\varepsilon_{jkab}\!+\!\Theta U_{[a}V_{b]}
\end{eqnarray}
so that
\begin{eqnarray}
&U_{a}U^{a}\!=\!-V_{a}V^{a}\!=\!\Theta^{2}\!+\!\Phi^{2}
\label{norm}\\
&U_{a}V^{a}=0
\label{orthogonal}
\end{eqnarray}
showing that of all the bi-linear spinor quantities, the two antisymmetric tensors are given in terms of all others, as the vector and axial-vector norms are given in terms of the scalar and pseudo-scalar, so that we have that
\begin{eqnarray}
&u_{a}u^{a}\!=\!-v_{a}v^{a}\!=\!1\\
&u_{a}v^{a}=0
\end{eqnarray}
if we define the directions $u^{\nu}$ and $v^{\nu}$ as
\begin{eqnarray}
&V^{a}\!=\!2\phi^{2}v^{a}\\
&U^{a}\!=\!2\phi^{2}u^{a}
\end{eqnarray}
and norms
\begin{eqnarray}
&\Theta\!=\!2\phi^{2}\sin{\beta}\\
&\Phi\!=\!2\phi^{2}\cos{\beta}
\end{eqnarray}
as a spinor re-parametrization: another simplification is performed noticing that in general $U^{a}$ is time-like and so we may boost in the frame where its space part vanishes making the spinor field reduce to the following
\begin{eqnarray}
&\!\!\psi\!=\!\left(\!\begin{tabular}{c}
$\pm \cos{\omega}e^{\frac{i}{2}\beta}e^{-\frac{i}{2}\varsigma}$\\
$\pm \sin{\omega}e^{\frac{i}{2}\beta}e^{\frac{i}{2}\varsigma}$\\
$\cos{\omega}e^{-\frac{i}{2}\beta}e^{-\frac{i}{2}\varsigma}$\\
$\sin{\omega}e^{-\frac{i}{2}\beta}e^{\frac{i}{2}\varsigma}$
\end{tabular}\!\right)\!e^{i\alpha}\phi
\end{eqnarray}
then we may have $V^{a}$ rotate as to be aligned along the third axis so to have the spinor reduced to either 
\begin{eqnarray}
&\!\psi\!=\!\left(\!\begin{tabular}{c}
$\pm e^{\frac{i}{2}\beta}$\\
$0$\\
$e^{-\frac{i}{2}\beta}$\\
$0$
\end{tabular}\!\right)\!e^{i\alpha}\phi\ \ \ \ \mathrm{or}
\ \ \ \ \psi\!=\!\left(\!\begin{tabular}{c}
$0$\\
$\pm e^{\frac{i}{2}\beta}$\\
$0$\\
$e^{-\frac{i}{2}\beta}$
\end{tabular}\!\right)\!e^{i\alpha}\phi
\label{spinor}
\end{eqnarray}
according to whether it is respectively the positive or the negative eigen-state of the rotation around the third axis, and eventually we may employ this third axis rotation in order to have the overall phase shifted away so that only the fields $\phi$ and $\beta$ remain, which are known as the module and the Takabayashi angle respectively and they account for all degrees of freedom of the spinor field. Eventually, we indicate $\boldsymbol{\nabla}_{\mu}$ as the spinorial covariant derivative used to define a dynamical Lagrangian for the material field.

From \cite{Fabbri:2014dxa}, we see that the Lagrangian is given in terms of the constants $M$, $X$ and $k$ being the torsion mass, the torsion and the gravitational coupling, with $m$ being the spinor mass, and varying such a Lagrangian yields
\begin{eqnarray}
\nonumber
&R^{\rho\sigma}\!-\!\frac{1}{2}Rg^{\rho\sigma}\!-\!\Lambda g^{\rho\sigma}
\!=\!\frac{k}{2}[\frac{1}{4}F^{2}g^{\rho\sigma}
\!-\!F^{\rho\alpha}\!F^{\sigma}_{\phantom{\sigma}\alpha}+\\
\nonumber
&+\frac{1}{4}(\partial W)^{2}g^{\rho\sigma}
\!-\!(\partial W)^{\sigma\alpha}(\partial W)^{\rho}_{\phantom{\rho}\alpha}+\\
\nonumber
&+M^{2}(W^{\rho}W^{\sigma}\!-\!\frac{1}{2}W^{2}g^{\rho\sigma})+\\
\nonumber
&+\frac{i}{4}(\overline{\psi}\boldsymbol{\gamma}^{\rho}\boldsymbol{\nabla}^{\sigma}\psi
\!-\!\boldsymbol{\nabla}^{\sigma}\overline{\psi}\boldsymbol{\gamma}^{\rho}\psi
\!+\!\overline{\psi}\boldsymbol{\gamma}^{\sigma}\boldsymbol{\nabla}^{\rho}\psi
\!-\!\boldsymbol{\nabla}^{\rho}\overline{\psi}\boldsymbol{\gamma}^{\sigma}\psi)-\\
&-\frac{1}{2}X(W^{\sigma}\overline{\psi}\boldsymbol{\gamma}^{\rho}\boldsymbol{\pi}\psi
\!+\!W^{\rho}\overline{\psi}\boldsymbol{\gamma}^{\sigma}\boldsymbol{\pi}\psi)]
\label{metricfieldequations}
\end{eqnarray}
and 
\begin{eqnarray}
&\nabla_{\rho}(\partial W)^{\rho\mu}\!+\!M^{2}W^{\mu}
\!=\!X\overline{\psi}\boldsymbol{\gamma}^{\mu}\boldsymbol{\pi}\psi
\label{torsionfieldequations}
\end{eqnarray}
alongside to
\begin{eqnarray}
&\nabla_{\sigma}F^{\sigma\mu}\!=\!q\overline{\psi}\boldsymbol{\gamma}^{\mu}\psi
\label{gaugefieldequations}
\end{eqnarray}
as the geometric field equations, with
\begin{eqnarray}
&i\boldsymbol{\gamma}^{\mu}\boldsymbol{\nabla}_{\mu}\psi
\!-\!XW_{\sigma}\boldsymbol{\gamma}^{\sigma}\boldsymbol{\pi}\psi\!-\!m\psi\!=\!0
\label{spinorfieldequation}
\end{eqnarray}
as the matter field equations; by following the methods we discussed in \cite{Fabbri:2016laz}, we may plug (\ref{spinor}) into the spinor field equation to obtain in terms of the vector and axial-vector
\begin{eqnarray}
&K_{\mu}\!=\!2XW_{\mu}\!+\!\frac{1}{2}\varepsilon_{\mu\sigma\nu\iota}\Omega^{\sigma\nu\iota}
\!-\!2(qA^{\rho}\!+\!\nabla^{\rho}\alpha)u_{[\rho}v_{\mu]}\label{K}\\
&\!\!G_{\mu}\!=\!\Omega_{\mu a}^{\phantom{\mu a}a}
\!-\!2(qA^{\rho}\!+\!\nabla^{\rho}\alpha)u^{\nu}v^{\sigma}\varepsilon_{\mu\rho\nu\sigma}\label{G}
\end{eqnarray}
the equivalent pair of field equations
\begin{eqnarray}
&\nabla_{\mu}\ln{\phi^{2}}\!-\!G_{\mu}\!+\!v_{\mu}2m\sin{\beta}\!=\!0\label{F1}\\
&\nabla_{\mu}\beta\!-\!K_{\mu}\!+\!v_{\mu}2m\cos{\beta}\!=\!0\label{F2}
\end{eqnarray}
and if we were to employ the last rotation to remove the phase we would simply have to set $\alpha\!=\!0$ leaving only the fields $\phi$ and $\beta$ in the field equations giving their dynamics.

The advantage of employing the last rotation to have the overall phase shifted off is that in doing so we will be left only with the module and the Takabayashi angle and with the spinor in a form displaying only the fundamental degrees of freedom, but in some cases it may be best to employ a form that is less reduced, as we discussed in \cite{Fabbri:2016fxt}.

The last element we need to introduce is the definition of the energy density, given as the source of the gravitational field equations (\ref{metricfieldequations}), according to the expression
\begin{eqnarray}
\nonumber
&E^{\rho\sigma}\!=\!\frac{1}{4}F^{2}g^{\rho\sigma}
\!-\!F^{\rho\alpha}\!F^{\sigma}_{\phantom{\sigma}\alpha}+\\
\nonumber
&+\frac{1}{4}(\partial W)^{2}g^{\rho\sigma}
\!-\!(\partial W)^{\sigma\alpha}(\partial W)^{\rho}_{\phantom{\rho}\alpha}+\\
\nonumber
&+M^{2}(W^{\rho}W^{\sigma}\!-\!\frac{1}{2}W^{2}g^{\rho\sigma})+\\
\nonumber
&+\frac{i}{4}(\overline{\psi}\boldsymbol{\gamma}^{\rho}\boldsymbol{\nabla}^{\sigma}\psi
\!-\!\boldsymbol{\nabla}^{\sigma}\overline{\psi}\boldsymbol{\gamma}^{\rho}\psi
\!+\!\overline{\psi}\boldsymbol{\gamma}^{\sigma}\boldsymbol{\nabla}^{\rho}\psi
\!-\!\boldsymbol{\nabla}^{\rho}\overline{\psi}\boldsymbol{\gamma}^{\sigma}\psi)-\\
&-\frac{1}{2}X(W^{\sigma}\overline{\psi}\boldsymbol{\gamma}^{\rho}\boldsymbol{\pi}\psi
\!+\!W^{\rho}\overline{\psi}\boldsymbol{\gamma}^{\sigma}\boldsymbol{\pi}\psi)
\label{energydensity}
\end{eqnarray}
and recalling that with (\ref{torsionfieldequations}) and (\ref{gaugefieldequations}), and (\ref{spinorfieldequation}), we have
\begin{eqnarray}
&\nabla_{\rho}E^{\rho\sigma}\!=\!0
\label{cons}
\end{eqnarray}
as conservation law, and similarly we define the density
\begin{eqnarray}
&2\phi^{2}\!=\!\rho
\label{density}
\end{eqnarray}
and recalling that with (\ref{spinorfieldequation}), we obtain
\begin{eqnarray}
&\nabla_{\mu}(\rho u^{\mu})\!=\!0
\label{cont}
\end{eqnarray}
as continuity equation, valid in the most general instance.
\section{Geometrical assessment of energy conditions}
As it was discussed in \cite{Fabbri:2016msm}, the non-relativistic regime is an approximation that is mathematically implemented for spinors by requiring the smallness of the spatial part of the vector $u^{a}$ and of the Takabayashi angle, and since the vector $u^{a}$ is the velocity the Takabayashi angle must be related to an internal motion: this is the manifestation of the presence of spin in spinorial fields. We will see the importance of this circumstance in what is coming next.

We begin the geometrical assessment of the energy issue by first considering the spinor field in its most reduced form, obtained when also the third axis rotation is done to shift the phase away: when $\alpha\!=\!0$ gets implemented in the form of the spinor (\ref{spinor}), and the covariant derivatives are computed, gravitational field equations (\ref{metricfieldequations}) are
\begin{eqnarray}
\nonumber
&R^{\rho\sigma}\!-\!\frac{1}{2}Rg^{\rho\sigma}\!-\!\Lambda g^{\rho\sigma}
\!=\!\frac{k}{2}[\frac{1}{4}F^{2}g^{\rho\sigma}
\!-\!F^{\rho\alpha}\!F^{\sigma}_{\phantom{\sigma}\alpha}+\\
\nonumber
&+\frac{1}{4}(\partial W)^{2}g^{\rho\sigma}
\!-\!(\partial W)^{\sigma\alpha}(\partial W)^{\rho}_{\phantom{\rho}\alpha}+\\
\nonumber
&+M^{2}(W^{\rho}W^{\sigma}\!-\!\frac{1}{2}W^{2}g^{\rho\sigma})+\\
\nonumber
&+\frac{1}{8}\rho(\Omega_{ab}^{\phantom{ab}\rho}\varepsilon^{\sigma abk}v_{k}
\!+\!\Omega_{ab}^{\phantom{ab}\sigma}\varepsilon^{\rho abk}v_{k}-\\
\nonumber
&-\Omega_{ijk}\varepsilon^{ijk\sigma}v^{\rho}
\!-\!\Omega_{ijk}\varepsilon^{ijk\rho}v^{\sigma})-\\
\nonumber
&-\rho\frac{q}{2}(A^{\rho}u^{\sigma}\!+\!A^{\sigma}u^{\rho}+\\
&+A_{k}u^{[k}v^{\sigma]}v^{\rho}\!+\!A_{k}u^{[k}v^{\rho]}v^{\sigma})
\!-\!m\rho\cos{\beta}v^{\rho}v^{\sigma}]
\end{eqnarray}
whose time-time component is
\begin{eqnarray}
\nonumber
&R^{00}\!-\!\frac{1}{2}R\!-\!\Lambda
\!=\!\frac{k}{2}[\frac{1}{4}F^{2}\!-\!F^{0b}\!F^{0}_{\phantom{0}b}+\\
\nonumber
&+\frac{1}{4}(\partial W)^{2}\!-\!(\partial W)^{0b}(\partial W)^{0}_{\phantom{0}b}+\\
\nonumber
&+\frac{1}{2}M^{2}(|W^{0}|^{2}\!+\!\vec{W}\!\cdot\!\vec{W})+\\
&+\rho(\frac{1}{2}\Omega^{120}\!-\!qA^{0})]
\end{eqnarray}
showing that the energy density is given by a contribution of the gauge fields, which is positive, plus a contribution of torsion with mass, also positive, plus the contribution of the spinor given by a term representing the interaction with the electric potential and a term that represents the interaction with gravity and whose overall sign is totally undetermined; because there exist fields that are neutral but still have problems of non-defined energy, it is clear that such a problem cannot be solved in terms of a so far unaccounted electrodynamic interaction, and so we may as well neglect it without losing information but getting a clearer form of the field equations given by
\begin{eqnarray}
&R^{00}\!-\!\frac{1}{2}R\!-\!\Lambda
\!=\!\frac{k}{2}\frac{1}{2}\rho\Omega^{120}
\label{equation}
\end{eqnarray}
with energy density $\frac{1}{2}\rho\Omega^{120}$ and energy $\frac{1}{2}\Omega^{120}$ as in \cite{Fabbri:2016laz}.

In this expression, apart from the spinor density, there appears only the spin connection, and non-linearly in the curvatures of the left-hand side: this means that in order to tell what is the energy, one would have to exactly solve, for the spin connection, such a non-linear field equation.

In absence of exact solutions, it is still possible to tell that spinors with opposite sign of the energy correspond to space-times with opposite sign of the $\Omega_{120}$ component of the spin connection, but switching the sign of the spin connection in a non-linear equation in general is not going to give another solution, and two spinors having opposite energy in general are not simultaneously solutions of the entire system of field equations we have considered above.

In general, two spinors of opposite energy would find themselves in two totally different space-time geometries.

We have to clarify one point: the reason why the spinor energy turns out to be given by a contribution of the spin connection is that in deciding to reduce the spinor to the form given by (\ref{spinor}) we could do so only by employing the local active spinorial transformation and by doing this it is inevitable that the spin connection gets a non-inertial additional contribution: and in fact, the $\Omega^{12}_{\phantom{12}t}$ component is precisely the one that is produced with a local rotation around the third axis; we might think at this as the fact that we have reduced the spinors to the physical degrees of freedom by transferring their non-physical components into the non-inertial frame, in the same manner in which spinning tops can be studied at rest only by going into a rotating reference system. Albeit this generally holds for the third axis rotation, but also for the alignments with the third axis and the boosts into the rest frame, we may still assume that for the latter two transformations there be no considerable non-inertial effects and see what this is going to give us; if there are no non-inertial effects tied to these two transformations, then it is possible to reduce the spinor to the form (\ref{spinor}) without generating any term for the spin connection. This cannot be done for the third axis rotation and so we quite simply avoid to do it \cite{Fabbri:2016fxt}.

Having already seen that in presence of a gravitational field the energy problem may not even be a problem, we next try to go further and see what happens when gravity is not considered; the reason for this decision is that we want to parallel as much as possible the treatment that is followed in the usual approaches. Hence, another working hypothesis is that we will neglect the gravitational field.

With these working hypotheses, we may therefore employ spinors with form (\ref{spinor}) without any associated non-inertial effect; this means that the spin connection would account for gravity alone but since there is no gravity the spin connection will be equal to zero: in this case we may also find a global system of Galileian coordinates where we may write $\alpha\!=\!-P_{k}x^{k}$ with $P_{k}$ being constant and so from (\ref{K}, \ref{G}) plugged into (\ref{F1}, \ref{F2}) we obtain that
\begin{eqnarray}
&\!\!\nabla_{\mu}\ln{\phi^{2}}
\!+\!2(qA^{\rho}\!-\!P^{\rho})u^{\nu}v^{\sigma}\varepsilon_{\mu\rho\nu\sigma}
\!+\!v_{\mu}2m\sin{\beta}\!=\!0\\
&\!\!\nabla_{\mu}\beta\!-\!2XW_{\mu}\!+\!2(qA^{\rho}\!-\!P^{\rho})u_{[\rho}v_{\mu]}
\!+\!v_{\mu}2m\cos{\beta}\!=\!0
\end{eqnarray}
as one can straightforwardly see. From these we get that
\begin{eqnarray}
\nonumber
&P^{\nu}\!=\!m\cos{\beta}u^{\nu}\!+\!qA^{\nu}
\!+\!v^{[\nu}u^{\mu]}(\frac{1}{2}\nabla_{\mu}\beta\!-\!XW_{\mu})+\\
&+\frac{1}{2}\varepsilon^{\nu\rho\sigma\mu}v_{\rho}u_{\sigma}\nabla_{\mu}\ln{\phi^{2}}
\label{momentum}
\end{eqnarray}
giving the explicit form of the $P_{k}$ vector; we notice therefore that this vector called momentum is not an independent object but it is given in terms of the physical degrees of freedom of the spinor. Only if $\nabla_{\mu}\phi\!=\!\beta\!=\!0$ and without torsion it becomes $P^{i}\!=\!mu^{i}\!+\!qA^{i}$ as in the usual case.

The form (\ref{momentum}) is just the usual Gordon decomposition for the spinor in form (\ref{spinor}) showing that the momentum has a contribution due to the velocity and contributions due to the spin: the last ones are given by the non-trivial Takabayashi angle and the non-constant density field.

When the spinorial covariant derivative of (\ref{spinor}) is used together with expression (\ref{momentum}) in (\ref{energydensity}) we obtain that
\begin{eqnarray}
\nonumber
&E^{\rho\sigma}\!=\!\frac{1}{4}F^{2}g^{\rho\sigma}
\!-\!F^{\rho\alpha}\!F^{\sigma}_{\phantom{\sigma}\alpha}+\\
\nonumber
&+\frac{1}{4}(\partial W)^{2}g^{\rho\sigma}
\!-\!(\partial W)^{\sigma\alpha}(\partial W)^{\rho}_{\phantom{\rho}\alpha}+\\
\nonumber
&+M^{2}(W^{\rho}W^{\sigma}\!-\!\frac{1}{2}W^{2}g^{\rho\sigma})-\\
\nonumber
&-\frac{1}{2}\rho XW_{\mu}(v^{[\sigma}u^{\mu]}u^{\rho}\!+\!v^{[\rho}u^{\mu]}u^{\sigma}
\!+\!g^{\mu\sigma}v^{\rho}\!+\!g^{\mu\rho}v^{\sigma})+\\
\nonumber
&+\frac{1}{4}\rho\nabla_{\mu}\beta
(v^{[\sigma}u^{\mu]}u^{\rho}\!+\!v^{[\rho}u^{\mu]}u^{\sigma}
\!+\!g^{\mu\sigma}v^{\rho}\!+\!g^{\mu\rho}v^{\sigma})+\\
\nonumber
&+\frac{1}{4}\nabla_{\mu}\rho(\varepsilon^{\sigma\eta\alpha\mu}v_{\eta}u_{\alpha}u^{\rho}
\!+\!\varepsilon^{\rho\eta\alpha\mu}v_{\eta}u_{\alpha}u^{\sigma})+\\
&+\rho m\cos{\beta}u^{\sigma}u^{\rho}
\label{energydensityspinor}
\end{eqnarray}
as the total energy density; again, this is given by the contribution of the gauge fields plus a contribution of torsion with mass plus the contribution of the spinor given by a term representing the interaction with torsion and terms representing the dynamics of the spinor. Only if we have that $\nabla_{\mu}\rho\!=\!\beta\!=\!0$ and no interactions such an expression becomes $E^{\rho\sigma}\!=\!\rho m u^{\sigma}u^{\rho}$ as we have in the usual case.

The form (\ref{energydensityspinor}) shows that the energy of the spinor field has a contribution due to the velocity and contributions due to the spin given by the non-trivial Takabayashi angle and the non-constant density field as now expected.

The expressions for the pure spinorial contribution are
\begin{eqnarray}
&\!\!\!\!\!\!\!\!P^{\nu}\!=\!m\cos{\beta}u^{\nu}\!+\!\frac{1}{2}v^{[\nu}u^{\mu]}\nabla_{\mu}\beta
\!+\!\frac{1}{2}\varepsilon^{\nu\rho\sigma\mu}v_{\rho}u_{\sigma}\!\nabla_{\mu}\ln{\phi^{2}}
\end{eqnarray}
and
\begin{eqnarray}
\nonumber
&\!\!\!\!E^{\rho\sigma}\!=\!\frac{1}{4}\rho\nabla_{\mu}\beta
(v^{[\sigma}u^{\mu]}u^{\rho}\!+\!v^{[\rho}u^{\mu]}u^{\sigma}
\!+\!g^{\mu\sigma}v^{\rho}\!+\!g^{\mu\rho}v^{\sigma})+\\
\nonumber
&+\frac{1}{4}\nabla_{\mu}\rho(\varepsilon^{\sigma\eta\alpha\mu}v_{\eta}u_{\alpha}u^{\rho}
\!+\!\varepsilon^{\rho\eta\alpha\mu}v_{\eta}u_{\alpha}u^{\sigma})+\\
&+\rho m\cos{\beta}u^{\sigma}u^{\rho}
\end{eqnarray}
whose temporal component
\begin{eqnarray}
&P^{0}\!=\!E^{00}/\rho\!=\!m\cos{\beta}\!-\!\frac{1}{2}v^{\mu}\nabla_{\mu}\beta
\end{eqnarray}
would give that the energy is positive whenever
\begin{eqnarray}
&2m\cos{\beta}\!-\!v^{\mu}\nabla_{\mu}\beta>0
\label{condition}
\end{eqnarray}
and although this condition is not trivial, in the common treatment of quantum fields where the Takabayashi angle is systematically set to zero it does become trivial and it directly shows that the energy is positive defined.

To conclude, let us consider what happens if the angle of Takabayashi is neglected together with torsion: imposing the condition (\ref{cons}) it is possible to work out that
\begin{eqnarray}
\nonumber
&\rho mu^{\nu}\nabla_{\nu}u^{\sigma}\!=\!-\frac{1}{4}\nabla_{\nu}[\nabla_{\mu}\rho
(\varepsilon^{\sigma\eta\alpha\mu}v_{\eta}u_{\alpha}u^{\nu}+\\
&+\varepsilon^{\nu\eta\alpha\mu}v_{\eta}u_{\alpha}u^{\sigma})]\!+\!\rho qF^{\sigma\alpha}u_{\alpha}
\end{eqnarray}
as the equation for the matter distribution coupled to an external electro-dynamic field. With no external Lorentz force this equation is still not the Newton law because of the non-trivial contribution of the density field and it is only when $\nabla_{\mu}\rho\!=\!0$ that $u^{\nu}\nabla_{\nu}u^{\sigma}\!=\!0$ is finally recovered.

Condition $\nabla_{\mu}\rho\!=\!0$ tells that the density is everywhere zero except in a localized region where it is constant, and we see this as what gives the macroscopic approximation.

Therefore conditions $\nabla_{\mu}\rho\!=\!\beta\!=\!0$ have to be regarded as the approximations for which, by neglecting the angle of Takabayashi, we neglect the internal dynamics, and for which, by assuming the density to be constant, we assume the configuration that is also valid in macroscopic cases.

When both conditions are enforced, we obtain the limit known in quantum mechanics as Ehrenfest theorem.
\section{Conclusion}
In this paper, we passed under scrutiny the possibility that for spinor fields the energy could be problematic.

We began by noticing that no treatment of the energy can be complete unless the energy-sourced gravitational field equations are considered, and we have immediately seen that in this case no study can be conclusive until we solve the field equation (\ref{equation}) exactly; nevertheless, we did discuss how two spinors having opposite energy possess different space-time structures, so positive and negative energy spinors cannot be two simultaneous solutions for the entire system of field equations in general cases.

Therefore, there is not much we may say when the full coupling to gravity is considered, and so to extract some information we decided to parallel the usual treatment of quantum field theory neglecting the gravitational field, as well as other interactions; for the free spinor, we obtained the expression of the energy and we saw that in general it should indeed be reasonable that its sign be not defined, since in it we cannot distinguish the contribution due to the kinetic energy being positive and those contributions due to the chiral internal dynamics and binding potential being negative: however positive energy is ensured when the negative energy of the internal dynamics is concealed inside the matter distribution, and indeed (\ref{condition}) showed us that when the Takabayashi angle is neglected, in spinless configurations, only positive energies appear eventually.

According to whether gravity is considered or not, and whether internal dynamics and interactions are taken or not, there are different responses that can be given to the problem of energy, but none of them is a definitive claim that there is indeed a problem with energy.

\end{document}